\documentclass[aps,prl,twocolumn,groupedaddress]{revtex4-1}

\usepackage{amsmath}
\usepackage{graphicx}
\usepackage{braket}

\begin{document}



\title{Integrated nanoplasmonic quantum interfaces for room temperature single photon sources}



\author{Fr\'ed\'eric Peyskens}
\email[]{fpeysken@mit.edu}
\affiliation{Quantum Photonics Group, RLE, Massachusetts Institute of Technology, Cambridge, Massachusetts 02139, USA}
\author{Darrick Chang}
\affiliation{ICFO-Institut de Ciencies Fotoniques, The Barcelona Institute of Science and Technology, 08860 Castelldefels, Barcelona, Spain}
\author{Dirk Englund}
\affiliation{Quantum Photonics Group, RLE, Massachusetts Institute of Technology, Cambridge, Massachusetts 02139, USA}


\date{\today}

\begin{abstract}
We describe a general analytical framework of a nanoplasmonic cavity-emitter system interacting with a dielectric photonic waveguide. Taking into account emitter quenching and dephasing, our model directly reveals the single photon extraction efficiency, $\eta$, as well as the indistinguishability, $I$, of photons coupled into the waveguide mode. Rather than minimizing the cavity modal volume, our analysis predicts an optimum modal volume to maximize $\eta$ that balances waveguide coupling and spontaneous emission rate enhancement. Surprisingly, our model predicts that near-unity indistinguishability is possible, but this requires a much smaller modal volume, implying a fundamental performance trade-off between high $\eta$ and $I$ at room temperature. Finally, we show that maximizing $\eta I$ requires that the system has to be driven in the weak coupling regime because quenching effects and decreased waveguide coupling drastically reduce $\eta$ in the strong coupling regime.
\end{abstract}

\pacs{}

\maketitle


\section{Introduction}

Atomic and photonic quantum systems are central in many areas of quantum information processing, including quantum computing, communication, and precision sensing. \cite{refN1,ref2,ref3} A central remaining challenge is to improve the naturally weak interaction between single photons and single emitters. \cite{ref4} To this end, a plethora of approaches using dielectric (Ref. \cite{ref8,ref5,refE3,refE2,refE1,ref26,ref9,ref10}) as well as plasmonic  (Ref.\cite{ref6,refN2,ref11,ref12,ref23,ref24,ref25,refaX1}) cavities and waveguides has been suggested. While several theoretical studies analyzed the interaction between quantum emitters and dielectric (Ref. \cite{ref4,refN3,refN6,refN4,refN5}) or plasmonic (Ref. \cite{ref6,refN2,ref11}) waveguides, it has been a remaining issue to develop a comprehensive physical model of a nanoplasmonic cavity interacting with a single quantum emitter and evanescently coupled to a dielectric waveguide. \cite{ref28,refN9} 
In particular, there is a need for a comprehensive theoretical model to analyze the single photon extraction efficiency and indistinguishability of such integrated nanoplasmonic systems. 
\par
In this paper we present for the first time a general theory framework of an integrated nanoplasmonic quantum interface which incorporates the impact of quenching and dephasing on the single photon extraction efficiency $\eta$ and indistinguishability $I$. Our analysis yields optimal operating conditions to maximize $\eta$ and $I$ and gives clear physical intuition in the fundamental performance trade-offs. We reveal that $\eta$ is maximized for an optimum cavity modal volume $V_{\eta}^{\footnotesize \mbox{opt}}$ which is inversely proportional to the cavity $Q-$factor. On the other hand, $I$ can only be maximized for much smaller $V_{c}\ll V_{\eta}^{\footnotesize \mbox{opt}}$ at room temperature, imposing a fundamental limit on the $\eta I$ product. Finally, it is shown that the maximum $\eta I$ product is obtained for weak coupling because quenching effects and reduced waveguide coupling induce a huge decrease of $\eta$ in the strong coupling regime.


\section{Model}
\begin{figure}[ht]
\includegraphics[width=0.48\textwidth]{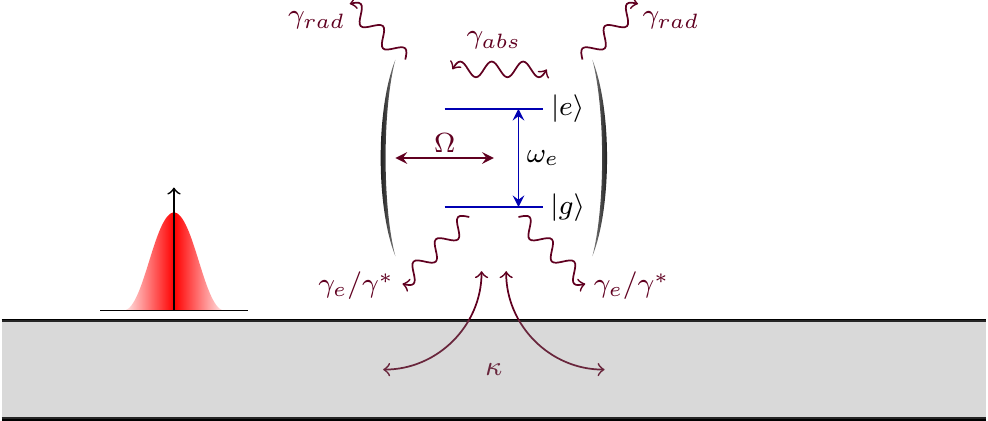}
\caption{\textbf{Quantum photonic platform.} The waveguide (gray) supports a 1D continuum of modes and interacts evanescently with a cavity-emitter system. All coupling rates and frequencies are explained in the main text. A short excitation pulse (red) through the waveguide is used to initialize the system into $\ket{e,0}$. \label{FigurePlatform}}
\end{figure}

The quantum photonic platform under investigation is shown in Fig. \ref{FigurePlatform}. It consists of a dielectric nanophotonic waveguide that evanescently interacts with a cavity-emitter system. The cavity has a resonance frequency $\omega_{c}$ and an overall linewidth $\gamma_{p}=\gamma_{c}+\kappa$, including the intrinsic linewidth $\gamma_{c}$ and the coupling rate $\kappa$ between the cavity mode (characterized by the operator $p$, annihilating a cavity excitation) and the waveguide mode; $\gamma_{c}=\omega_{c}/(2Q)=\gamma_{rad}+\gamma_{abs}$ with $\gamma_{rad}$ the radiative decay rate to the non-guided modes, $\gamma_{abs}$ the absorption decay rate and $Q$ the unloaded quality factor. The two-level quantum emitter has a resonance frequency $\omega_{e}$ between its ground $\ket{g}$ and excited $\ket{e}$ state ($S_{z}=1/2\left(\ket{e}\bra{e}-\ket{g}\bra{g}\right)$, $S_{+}=\ket{e}\bra{g}$, $S_{-}=\ket{g}\bra{e}$) and a decay rate $\gamma_{e}$. The emitter and cavity couple at a rate $\Omega$. We also allow for pure dephasing at a rate $\gamma^{*}$, which at room temperature can be orders of magnitude larger than $\gamma_{e}$.  \cite{refN7}
\par
We model the combined dissipative quantum system using a master equation in a frame rotating at $\omega_{e}$  (see Supplemental Material Ref. \cite{refSM})
\begin{multline}
\frac{d\rho}{dt}= -\frac{i}{\hbar}[\mathcal{H}_{rot},\rho] + \frac{\gamma_{p}}{2}\left(2 p\rho p^{\dagger} - p^{\dagger}p\rho - \rho p^{\dagger}p \right) \\ +\frac{\gamma_{e}}{2}\left(2 S_{-}\rho S_{+} - S_{+}S_{-}\rho - \rho S_{+}S_{-} \right) \\ +\frac{\gamma^{*}}{2}\left(2 S_{z}\rho S_{z} - S_{z}S_{z}\rho - \rho S_{z}S_{z} \right),
\end{multline}
where $\mathcal{H}_{rot}=\hbar\delta_{c}p^{\dagger}p+\hbar \Omega \left(pS_{+}+p^{\dagger}S_{-}\right)$ with $\delta_{c}=\omega_{c}-\omega_{e}$. We assume that a short excitation pulse initializes the atomic emitter into the excited state $\ket{e}$ and subsequently observe the coupling back into the waveguide. The system dynamics can then be described in a Hilbert space consisting of 3 states: $\{\ket{g,0},\ket{g,1},\ket{e,0}\}$ respectively corresponding to the atom in the ground state and no photon in the cavity, the atom in the ground state and one photon in the cavity or the atom in the excited state and no photon in the cavity. The photon can leave the cavity either by decaying radiatively to the non-guided or guided modes, or non-radiatively by cavity absorption.
\par
For dielectric cavities the emitter decay rate $\gamma_{e}$ is approximated by $\gamma_{d}=\sqrt{\epsilon_{d}}\gamma_{0}$ with $\gamma_{0}=\frac{8\pi^{2}}{3\hbar\epsilon_{0}}\frac{|\textbf{d}|^{2}}{\lambda^{3}}$ the vacuum decay rate of a dipole emitter, $\textbf{d}$ the dipole moment vector, and $\epsilon_{d}$ the relative permittivity of the background dielectric. For a nanoplasmonic cavity this assumption is invalid due to quenching effects near the metal surface. To be specific, we assume here a spherical metallic nanoparticle with radius $R$ and quality factor $Q=15$ to model the nanoplasmonic cavity, as $\Omega$ and $\gamma_{e}$ can be calculated analytically for this specific case. The decay rate $\gamma_{e}$ for an emitter placed at a distance $d$ ($\xi=d/R$) from the metal surface can be approximated as the sum of the radiative decay rate and the quenching rate, $\gamma_{e}\approx \gamma_{d}+\Omega^{2}f_{q}$ where
\begin{equation}
\footnotesize
f_{q}=\sum_{l=2}^{\infty} \left(\frac{(2l+1)(l+1)^{2}}{12l (1+\xi)^{2l-2}}\right)\left(\frac{\gamma_{c}}{\left(\omega_{e}-\omega_{l}\right)^{2}+\left(\frac{\gamma_{c}}{2}\right)^{2}}\right)
\end{equation}
\normalsize
the quenching factor, $\omega_{l}$ the resonance frequency of the plasmon modes and $\Omega=\sqrt{\frac{9\pi c^{3}\gamma_{0}}{2\omega_{p}^{2}\epsilon_{d}V_{c}(1+\xi)^{6}}}$ the coupling constant to the plasmonic dipole mode, with $V_{c}=\frac{\pi R^{3}}{\epsilon_{d}}$ the cavity modal volume and $\omega_{p}$ the plasma frequency (see Ref. \cite{refSM} for the full derivation). We assumed here that the emitter is on resonance with the fundamental plasmonic mode, i.e. $\omega_{c}=\omega_{1}=\omega_{e}$. This effective decay rate $\gamma_{e}$ takes the quenching due to higher order plasmon modes into account.

\section{On-chip single photon generation}

\subsection{Efficiency}

The efficiency of emitting a single photon into the waveguide, $\eta$, is calculated from the time-averaged number of intracavity photons multiplied by their decay rate into the waveguide, i.e. $\eta=\kappa\int_{0}^{\infty}dt\langle p^{\dagger}(t)p(t)\rangle=\kappa\int_{0}^{\infty}dt\langle g,1\left|\rho(t)\right|g,1\rangle$. The solution of the time-dependent master equation, with $\delta_{c}=0$ and the initial state of the system in $\ket{e,0}$, can be found analytically,
\begin{equation}
\eta\approx\frac{\kappa}{(\gamma_{e}+\gamma_{p})\left(1+\frac{\gamma_{e}\gamma_{p}}{4\Omega^{2}}\right)},
\end{equation}
and is independent of the dephasing rate $\gamma^{*}$ since $\gamma_{p}\gg\gamma^{*}$ for a nanoplasmonic cavity (see Ref. \cite{refSM}). The coupling constant between the cavity and the waveguide modes $g_{\footnotesize \mbox{wg}}$ is proportional to the cavity dipole moment, which itself is proportional to $V_{c}$, and inversely proportional to $\sqrt{V_{c}}$ due to energy normalization. As such, the decay rate $\kappa=\frac{4\pi g_{\footnotesize \mbox{wg}}^{2}}{c}$ will increase linearly with $V_{c}$ and in general is given by
\begin{equation}
\kappa=\frac{\omega_{c}^{4}\chi^{\kappa}\alpha_{0}^{2}\epsilon_{\footnotesize \mbox{eff}}}{2\pi^{2}c^{3}\epsilon_{d}\epsilon_{\footnotesize \mbox{wg}}}V_{c},
\end{equation}
with $\epsilon_{\footnotesize \mbox{eff}}$ the relative effective mode permittivity (defining the effective modal area of the waveguide mode $A_{\footnotesize \mbox{eff}}=\lambda_{c}^{2}/(4\epsilon_{\footnotesize \mbox{eff}})$), $\epsilon_{\footnotesize \mbox{wg}}$ the relative permittivity of the waveguide core, $\chi^{\kappa}$ a factor incorporating the overlap between the waveguide and the cavity mode ($0<\chi^{\kappa}\leq 1$) and $\alpha_{0}$ a constant determining the polarizability $\alpha_{p}$ of the cavity mode, $\alpha_{p}=\epsilon_{0}\alpha_{0}V_{c}$ (see Ref. \cite{refSM}). For a spherical particle, $\alpha_{0}=2\epsilon_{d}$, such that $\kappa=\frac{2R^{3}\chi^{\kappa}\omega_{c}^{4}\epsilon_{\footnotesize \mbox{eff}}}{\pi c^{3}\epsilon_{\footnotesize \mbox{wg}}}$. 
\begin{figure*}[ht]
\includegraphics[width=\textwidth]{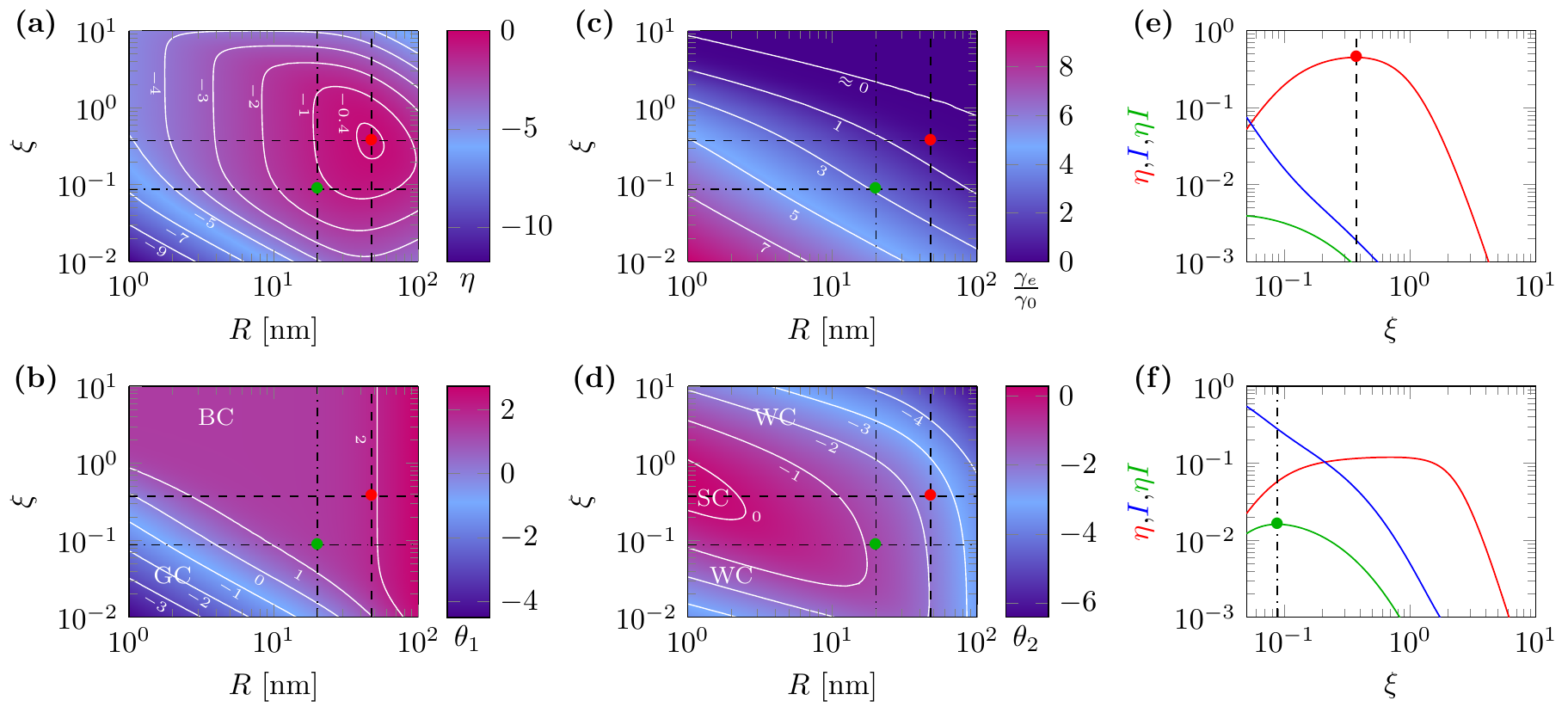}
\caption{Single photon characteristics ($\lambda_{c}=637$ nm) for a nanoplasmonic cavity with $Q=15$, $\gamma_{d}=1$ GHz, $\gamma^{*}=3.5$ THz (a realistic value at room temperature, Ref. \cite{refN7}), $\epsilon_{\footnotesize \mbox{eff}}=4$, $\epsilon_{\footnotesize \mbox{wg}}=4$, $\chi^{\kappa}=1$ and the incorporation of 1000 higher-order modes. \textbf{(a)} Single photon extraction efficiency $\eta$. \textbf{(b)} $\theta_{1}=\gamma_{p}/(\gamma_{e}+\gamma^{*})$ defining the bad ($\theta_{1}\gg 1$, BC) and good ($\theta_{1}\ll 1$, GC) cavity limit. \textbf{(c)} $\gamma_{e}/\gamma_{0}$ ratio to assess the quenching strength. \textbf{(d)} $\theta_{2}=2\Omega/(\gamma_{e}+\gamma_{p}+\gamma^{*})$ defining the regions of strong ($\theta_{2}\gg 1$, SC) and weak ($\theta_{2}\ll 1$, WC) coupling. All surface plots are on a $\log_{10}$ scale. \textbf{(e-f)} $\eta$ (red), $I$ (blue) and $\eta I$ (green) for two $R$ slices: \textbf{(e)} $R=20$ nm (vertical dashed line in Figs. \textbf{(a-d)}) and \textbf{(f)} $R=47.5$ nm (vertical dashdotted line in Figs. \textbf{(a-d)}). The intersection between either the vertical and horizontal dashed (red circle) or dashdotted (green circle) lines in Figs. \textbf{(a-d)} respectively marks the maximum of $\eta$ in \textbf{(e)} and the maximum of $\eta I$ in \textbf{(f)}. \label{FigureSPGE}}
\end{figure*}
\par
The analytical formulas for $\kappa$, $\Omega$ and $\gamma_{e}$ now allow us to evaluate $\eta$ as a function of $\xi$ and $R$. As seen in Fig. \ref{FigureSPGE}(a), there exists an optimum $\xi$ and $R$ that maximizes $\eta$ (marked by the red dot). This optimum can be understood from the limiting cases for $\eta$. Figure \ref{FigureSPGE}(b) plots the ratio between the loaded cavity decay rate $\gamma_{p}$ and the total emitter linewidth, which we define here as $\theta_{1}=\gamma_{p}/(\gamma_{e}+\gamma^{*})$. $\theta_1$ distinguishes the bad cavity regime ($\theta_{1}\gg 1$) from the good cavity regime ($\theta_{1}\ll 1$), following Ref. \cite{refN7}. Comparing Fig. \ref{FigureSPGE}(a) and Fig. \ref{FigureSPGE}(b) shows that the parameters for optimal $\eta$ clearly fall into the bad cavity limit for which $\eta\approx\left(\left(1+\frac{\gamma_{c}}{\kappa}\right)\left(1+\frac{\gamma_{e}(\kappa+\gamma_{c})}{4\Omega^{2}}\right)\right)^{-1}$. Since $\gamma_{e}=\gamma_{d}+\Omega^{2}f_{q}$ is an explicit function of $\Omega$, two limiting cases emerge. Either $\eta$ is limited by the intrinsic decay rate, i.e. $\gamma_{e}\approx\gamma_{d}$, or by quenching, i.e., $\gamma_{e}\approx \Omega^{2}f_{q}$ (see the plot of $\gamma_{e}/\gamma_{d}$ in Fig. \ref{FigureSPGE}(c)). For large $\xi>10$, the quenching factor $f_{q}\propto \xi^{-2}$ approaches zero and any increase in $\xi$ only reduces $\Omega$, implying a concomitant decrease in $\eta$. On the other hand, if $\xi\rightarrow 0$, $f_{q}\propto \xi^{-3}$, eventually reducing the single photon emission as well. As such the optimum $\xi_{opt}$ is expected at the cross-over region between the intrinsic decay rate limited and quenching limited case (which is confirmed by Fig. \ref{FigureSPGE}(c)). The optimum in $R$ strikes a balance between an increased cavity-emitter coupling strength $\Omega^2\propto 1/V_{c}$ due to the confinement of the cavity photon to a small modal volume, and a simultaneous reduction in the radiative emission rate $\kappa \propto V_c$ due to the smaller dipole moment. While one desires a large out-coupling to the waveguide mode $\kappa$ as large as possible, a continuous increase in $R$ ($V_{c}$) also reduces $\Omega$, eventually reducing $\eta$ again. In the Supplemental Material (Ref. \cite{refSM}) it is shown that for $\xi$ values in the cross-over region, the optimal cavity modal volume is approximately given by
\begin{equation}
V_{\eta}^{\footnotesize \footnotesize \mbox{opt}}\approx\frac{\lambda_{c}^{3}}{Q}\left(\frac{\epsilon_{d}\epsilon_{\footnotesize \mbox{wg}}}{8\pi\chi^{\kappa}\alpha_{0}^{2}\epsilon_{\footnotesize \mbox{eff}}}\right).
\end{equation}
The optimum cavity modal volume hence has to increase if either the $Q-$factor, the cavity polarizability or the overlap between the cavity and the waveguide mode is reduced. Similarly, if the effective modal area increases (i.e. lower $\epsilon_{\footnotesize \mbox{eff}}$), the waveguide mode becomes less confined and will spread out over a larger area in the vicinity of the waveguide core. Consequently, the evanescent coupling with nearby cavity-emitter systems will decrease, again necessitating an increase in $V_{c}$. This result clearly demonstrates that it is not always beneficial to achieve the smallest possible modal volume and hence serves as an important caveat to the commonly used strategy of reducing it to increase cavity-emitter interactions. In particular, while the interaction strength does increase for smaller $V_{c}$, such a strength is not helpful if the information contained in the cavity photon does not escape into the waveguide.

\subsection{Indistinguishability}
Apart from generating single photons with high efficiency, many applications require the photons to be indistinguishable. Therefore we also investigate the indistinguishability $I$ of two photons emitted by an integrated nanoplasmonic cavity. Depending on the previously defined ratio $\theta_{1}$ and the ratio $\theta_{2}=2\Omega/(\gamma_{e}+\gamma_{p}+\gamma^{*})$ (shown in Fig. \ref{FigureSPGE}(d)), one can distinguish three limiting cases for the indistinguishability, as introduced in Ref. \cite{refN7}. From Fig. \ref{FigureSPGE}(d) it is clear that $\theta_{2}\ll 1$ for most parameter values, meaning that our system is in the weak coupling regime. In this regime, one can either operate in the previously introduced good cavity or bad cavity limit. Achieving high $I$ in the bad cavity limit is difficult for dielectric cavities since their coupling strengths usually cannot compensate for room temperature dephasing. \cite{refN7} On the contrary, nanoplasmonic antennas allow much larger coupling strengths, turning the bad cavity limit into a regime of practical interest. In this limit, the effect of the cavity is to add an extra channel of irreversible emission at a rate $R=4\Omega^{2}/\left(\gamma_{e}+\gamma^{*}+\gamma_{p}\right)$, such that the dynamics of the coupled system can be described by an effective quantum emitter with decay rate $\gamma_{e}+R$, i.e. $I=\left(\gamma_{e}+R\right)/\left(\gamma_{e}+R+\gamma^{*}\right)$. \cite{refN7} Maximizing $I$ requires the cavity modal volume to be much smaller than an upper limit $V_{u}$,
\begin{equation}
V_{c}\ll V_{u}=V_{\eta}^{\footnotesize \mbox{opt}}\left(\frac{\alpha_{0}}{\epsilon_{d}}\right)^{2}\left(\frac{\epsilon_{\footnotesize \mbox{eff}}}{\epsilon_{\footnotesize \mbox{wg}}}\right)\left(\frac{\gamma_{0}}{\gamma^{*}}\right)Q^{2}\chi^{\kappa}.
\end{equation}
This constraint mainly stems from the fact that the cavity-emitter coupling must be large enough to exceed the dephasing rate, necessitating very small modal volumes. However, $V_{u}$ is typically much smaller than the optimum modal volume for $\eta$ as room temperature $\gamma_{0}/\gamma^{*}$ ratios are usually on the order of $10^{-3}$ or $10^{-4}$. Our model therefore shows conflicting requirements for maximizing $I$ and $\eta$ at room temperature. This trade-off between $\eta$ and $I$ is shown in Figs. \ref{FigureSPGE}(e-f) where $\eta$, $I$ and $\eta I$ are evaluated along the $\xi-$slice for which $\eta$ (e) or $\eta I$ (f) are maximized respectively. The $(\xi,R)$ combination that maximizes $\eta I$ is marked by a green dot in Figs. \ref{FigureSPGE}(a-d). Figure \ref{FigureSPGE}(e) shows that for the given parameters, $\eta\approx 45\%$ at $\xi\approx 0.38$ and $R\approx 47.5$ nm. Nevertheless the indistinguishability is low ($\approx0.2\%$). When considering the $R-$slice along which $\eta I$ is maximal ($\approx 1.6\%$ for $R=20$ nm and $\xi=0.09$), $\eta$ drops to $\sim 5.7\%$, while $I$ increases to $\sim 28\%$ (Fig. \ref{FigureSPGE}(f)). Indeed, the increase in $I$ is accompanied by a $(47.5/20)^{3}\approx 10-$fold reduction in cavity modal volume. For sufficiently small $\xi$ and/or $R$ values, one can moreover achieve the good cavity or strong coupling limit (see Fig. \ref{FigureSPGE}(b) and (d)). In both regimes it is possible to generate photons with almost perfect indistinguishability at room temperature. However, the reduction in $\xi$ and $R$ induce large quenching effects (Fig. \ref{FigureSPGE}(c)), which eventually sharply reduce the single photon extraction efficiency (see Fig. \ref{FigureSPGE}(a) for $\eta$ values in the good cavity or strong coupling region); for example: $I\approx 93\%$ but $\eta\approx 0.001\%$ for $R\approx 4$ nm and $\xi\approx 0.035$. The reduction in $R$ moreover lowers $\eta$ through $\kappa$. So in any of the three limiting cases there will be a trade-off between high $I$ and high $\eta$. For dielectric cavities, high $I$ and high $\eta$ can only be achieved simultaneously in the strong coupling regime. \cite{refN7} By contrast, we have shown that the maximum $\eta I$ performance is achieved in the bad cavity limit. This is an important observation since most recent reports focus on achieving strong coupling (Ref. \cite{ref29,refN10,refSNJ1}). While the strong coupling regime is beneficial for obtaining high $I$, it is also accompanied by a strong reduction in $\eta$ due to quenching and reduced waveguide coupling.
\par
It is however possible to raise $V_{u}$, to partially reconcile the different constraints on the modal volume in the bad cavity limit. Apart from maximizing the overlap between the cavity and the waveguide mode ($\chi^{\kappa}\rightarrow 1$), there are two other ways to improve the system. First, the waveguide geometry should be optimized to maximize the effective permittivity $\epsilon_{\footnotesize \mbox{eff}}$ ($\leq\epsilon_{\footnotesize \mbox{wg}}$). Second, the polarizability of the fundamental dipole mode should be increased as much as possible (i.e. $\alpha_{0}\gg\epsilon_{d}$), for example by considering rod or bowtie antennas (Ref. \cite{ref22,ref21}). Engineered antenna geometries can also achieve larger field enhancements as compared to the spherical particle case, which in turn result in higher $\eta$ or $\eta I$. In the Supplemental Material (Ref. \cite{refSM}) it is shown that an enhancement in the field and the polarizability by a factor 10 already allows single photon extraction efficiencies of $80\%$ while $\eta I$ increases to $50\%$.
Further optimization could enable room temperature on-demand ($\eta\rightarrow 1$) or perfectly indistinguishable ($I\rightarrow 1$) single photon sources. However, achieving both at the same time (i.e. $\eta I=1$) appears impossible for any realistic on-chip nanoplasmonic quantum interface at room temperature due to the conflicting requirements on the modal volume. If the system is cooled down such that the dephasing rate $\gamma^{*}$ drops, it is possible to achieve much higher $\eta I$; a reduction of the dephasing rate to $\mathcal{O}(\text{GHz})$, while keeping all other parameters the same as in Figure \ref{FigureSPGE}, allows $\eta I\approx 35\%$. In the absence of dephasing, $\eta I$ eventually becomes limited by $\eta$. Nevertheless it should be noted that the obtained $\eta I=1.6\%$ at room temperature (assuming a perfect overlap between the cavity and the waveguide mode) is already orders of magnitude higher than what can be achieved using spectral filtering and moreover approaches earlier reported values in the good cavity regime. \cite{refN7}

\section{Conclusion}

We presented a theoretical framework for a nanoplasmonic cavity-emitter system evanescently coupled to a dielectric waveguide and investigated the impact of quenching and dephasing on the single photon extraction efficiency $\eta$ and indistinguishability $I$. While quenching imposes fundamental limits on both $\eta$ and $I$, dephasing only has a considerable impact on $I$. We showed there exists an optimum cavity modal volume $V_{\eta}^{\footnotesize \mbox{opt}}$ to maximize $\eta$, which balances the coupling rate into the waveguide with the emission rate enhancement. This optimum is inversely proportional to the cavity $Q-$factor and depends on the effective modal area of the waveguide and cavity polarizability. Unfortunately, $V_{\eta}^{\footnotesize \mbox{opt}}$ is typically much larger than the modal volume required to maximize $I$. This trade-off imposes a fundamental limit on the $\eta I$ product and inhibits the perfect $\eta I=1$ value to be obtained for most experimentally achievable nanoplasmonic systems at room temperature. While the strong coupling regime is beneficial for achieving high $I$, we moreover showed that the maximal $\eta I$ is obtained for weak coupling. Furthermore we addressed strategies, such as field and cavity polarizability enhancement, to partially alleviate the $\eta I$ trade-off. Our theoretical framework captures the essential physics of integrated nanoplasmonic cavities for quantum photonic applications and identifies all the important design parameters to guide future efforts in the development of integrated single photon sources at low and room temperature.

\section{Acknowledgements}
F.P. acknowledges support from a BAEF (Belgian American Educational Foundation) and Fulbright postdoctoral fellowship. D.E.C. acknowledges support from Fundacio Privada Cellex, Spanish MINECO Severo Ochoa Programme SEV-2015-0522, MINECO Plan Nacional Grant CANS, CERCA Programme/Generalitat de Catalunya, and ERC Starting Grant FOQAL.

\newpage

\section{Supplemental Material}

\subsection{Master equation}
The nanophotonic waveguide supports a 1D continuum of left and right traveling modes which are characterized by the operators $l_{k}$ and $r_{k}$ respectively ($l_{k}$ and $r_{k}$ annihilate a left- or right- traveling photon with wavenumber $k=\omega_{k}/c$). The Hamiltonian of this quantum photonic platform is then given by
\begin{multline}
\mathcal{H} =\hbar\omega_{e}S_{z} + \hbar\omega_{c}p^{\dagger}p + \hbar \Omega \left(pS_{+}+p^{\dagger}S_{-}\right)  \\
+\hbar\int dk \omega_{k}l^{\dagger}_{k}l_{k} + \hbar \int dk \omega_{k}r^{\dagger}_{k}r_{k} + \\
+ \hbar g_{wg} \int dk\left(l^{\dagger}_{k}p + l_{k}p^{\dagger}\right) +\hbar g_{wg} \int dk\left(r^{\dagger}_{k}p + r_{k}p^{\dagger}\right).
\end{multline}
It includes the free Hamiltonian of both the emitter, the cavity and the waveguide modes as well as the interaction between the emitter and the cavity and the interaction between the cavity and waveguide modes. The coupling constant between the cavity and the waveguide modes is $g_{wg}=\sqrt{\frac{c\kappa}{4\pi}}$. \cite{ref20} Following an approach similar to References \cite{ref20,ref15,ref16} we can formally solve the Heisenberg equation for the $r_{k}$ and $l_{k}$ operators:
\begin{align}
	r_{k}(t)&=r_{k}(t_{0})\text{e}^{-i\omega_{k}(t-t_{0})}-ig_{wg}\int_{t_{0}}^{t}p(u)\text{e}^{-i\omega_{k}(t-u)}du \nonumber \\
	l_{k}(t)&=l_{k}(t_{0})\text{e}^{-i\omega_{k}(t-t_{0})}-ig_{wg}\int_{t_{0}}^{t}p(u)\text{e}^{-i\omega_{k}(t-u)}du \nonumber
\end{align}
\normalsize
When inserting these formal solutions into the solutions of the cavity operator, one finds
\begin{align}
\dot{p}&=-i\omega_{c}p(t)-ig_{wg}\left(\int r_{k} dk+\int l_{k} dk\right)- i\Omega S_{-} \nonumber \\
		&=-i\omega_{c}p(t)-ig_{wg}\left(-\frac{2i\pi g_{wg}}{c}p(t)\right)- i\Omega S_{-} \nonumber \\
		&=-i\omega_{c}p(t)-\frac{2\pi g_{wg}^{2}}{c}p(t)- i\Omega S_{-} \nonumber \\
		&=-i\omega_{c}p(t)-\frac{\kappa}{2}p(t)- i\Omega S_{-}. \label{eqp}
\end{align}
\normalsize
Similar to the results obtained in Reference \cite{ref20}, one can see from equation (\ref{eqp}) that the infinite waveguide degrees of freedom can be effectively integrated out such that the dynamics of the overall system can be accurately described by incorporating an additional Lindblad term to the master equation
\begin{equation}
 \sum_{\nu} \left(2 O_{\nu}\rho O_{\nu}^{\dagger} - O_{\nu}^{\dagger}O_{\nu}\rho - \rho O_{\nu}^{\dagger}O_{\nu}\right)
\end{equation}
with 
\begin{equation}
O_{\nu}= \sqrt{\frac{\kappa}{4}}p,\ \ \ \ \ \ \ \nu=\pm
\end{equation}
where $\nu$ distinguishes the right- and left-propagating fields. This additional Lindblad term
\begin{equation}
\frac{\kappa}{2}\left(2 p\rho p^{\dagger} - p^{\dagger}p\rho - \rho p^{\dagger}p\right)
\end{equation}
describes the decay of the cavity into the waveguide modes. When switching to a frame rotating at $\omega_{L}$ and adding the extra Lindblad term, the master equation eventually becomes
\begin{multline}
\frac{d\rho}{dt}= -\frac{i}{\hbar}[\mathcal{H}_{rot},\rho] + \frac{\gamma_{p}}{2}\left(2 p\rho p^{\dagger} - p^{\dagger}p\rho - \rho p^{\dagger}p \right) \\ +\frac{\gamma_{e}}{2}\left(2 S_{-}\rho S_{+} - S_{+}S_{-}\rho - \rho S_{+}S_{-} \right) \\ +\frac{\gamma^{*}}{2}\left(2 S_{z}\rho S_{z} - S_{z}S_{z}\rho - \rho S_{z}S_{z} \right)
\end{multline}
where
\begin{equation}
\mathcal{H}_{rot}=\hat{U}\mathcal{H}\hat{U}^{\dagger}-\hat{A}=\mathcal{H}_{atom}+\mathcal{H}_{cavity}+\mathcal{H}_{coupling}
\end{equation}
with
\begin{equation}
\hat{U}=\exp\left(-\frac{i\hat{A}t}{\hbar}\right),\ \ \ \ \ \ \ \hat{A}=\hbar\omega_{L}\left(p^{\dagger}p+S_{z}\right)
\end{equation}
and
\begin{align}
\mathcal{H}_{atom} &= \hbar\left(\omega_{e}-\omega_{L}\right)S_{z} = \hbar\delta_{e}S_{z} \\
\mathcal{H}_{cavity} &= \hbar\left(\omega_{c}-\omega_{L}\right)p^{\dagger}p = \hbar\delta_{c}p^{\dagger}p \\
\mathcal{H}_{coupling} &=  \hbar \Omega \left(pS_{+}+p^{\dagger}S_{-}\right)
\end{align}

\subsection{Decay rate near plasmonic nanoparticle}

In this section we will correlate the parameters as appearing in the quantum master equation (i.e. $\Omega,\gamma_{c},\gamma_{e}$) with the parameters obtained for dipole emission near a spherical metallic nanoparticle. This can be done in the weak coupling regime when the emitter is expected to decay exponentially and obeys Fermi's Golden Rule. For a spherical particle in the electrostatic limit one can obtain a simple expression for Fermi's Golden Rule by a multipole expansion of the coupling to the plasmon modes. We will show this allows to extrapolate $\Omega$ and $\gamma_{e}$ for near-resonant coupling with the fundamental dipole mode. Subsequently we discuss how the remaining radiative contribution to the emitter decay rate, which is not included in the electrostatic approximation, can be calculated.
\par
The overall decay rate of a quantum emitter $\gamma_{tot}$ is given by
\begin{equation}
\gamma_{tot}=\frac{6\pi c\gamma_{0}}{\omega}\textbf{u}_{d}\cdot \left[\Im (\textbf{G}(\textbf{r}_{e},\omega))\right] \cdot \textbf{u}_{d}
\end{equation}
where $\gamma_{0}=\frac{8\pi^{2}}{3\hbar\epsilon_{0}}\frac{|\textbf{d}|^{2}}{\lambda_{c}^{3}}$ is the decay rate of the emitter in vacuum, $\textbf{u}_{d}$ the unit vector along the dipole moment $|\textbf{d}|$ and $\textbf{G}(\textbf{r},\omega)$ the Green tensor associated to the electromagnetic environment of the dipole (evaluated at the emitter position $\textbf{r}_{e}$). \cite{ref25} This decay rate consists of a radiative and non-radiative contribution. For an emitter in a background medium with relative permittivity $\epsilon_{d}$ and near a nanoplasmonic cavity, the radiative part contains both free-space emission ($\approx\gamma_{d}=\gamma_{0}\sqrt{\epsilon_{d}}=\gamma_{0}n_{d}$) as well as cavity-mediated radiation ($\gamma_{rad}^{c}$), while the non-radiative part ($\gamma_{nrad}^{c}$) is attributed to metal absorption. As such
\begin{equation}
\gamma_{tot}=\gamma_{d}+\gamma_{rad}^{c}+\gamma_{nrad}^{c}.
\end{equation}
If the emitter is positioned at a distance $d$ from the spherical metallic surface (radius $R$) one can calculate $\gamma_{nrad}^{c}$ in the electrostatic approximation (Ref. \cite{ref25}),
\footnotesize
\begin{equation}
\gamma_{nrad}^{c}=\sum_{l=1}^{\infty} \Omega_{1}^{2}\left(\frac{(2l+1)(l+1)^{2}}{12l (1+\xi)^{2l-2}}\right)\left(\frac{\frac{\gamma_{i}}{2}}{\left(\omega-\omega_{l}\right)^{2}+\left(\frac{\gamma_{i}}{2}\right)^{2}}\right)
\end{equation}
\normalsize
with $\xi=d/R$,
\begin{equation}
\Omega_{1}^{2} = \frac{9c^{3}\gamma_{0}}{\omega_{p}^{2}R^{3}(1+\xi)^{6}}
\end{equation}
the coupling constant to the dipole mode ($l=1$) and
\begin{equation}
\omega_{l}=\omega_{p}\sqrt{\frac{l}{l\epsilon_{\infty}+(l+1)\epsilon_{d}}}
\end{equation}
the resonance frequencies of the plasmon modes. Moreover, $\omega_{p}$ is the plasma frequency of the metal and $\gamma_{i}$ the intrinsic metal absorption loss as determined by the Drude model
\begin{equation}
\epsilon_{M}=\epsilon_{\infty}-\frac{\omega_{p}^{2}}{\omega^{2}+i\omega\gamma_{i}}.
\end{equation}
If the spherical particle is small enough (i.e. $R\ll\lambda$), the dipolar mode ($l=1$) will be the only one with non-vanishing dipole moment. \cite{ref22} Therefore it is reasonable to assume that the dipole mode is the only one that will contribute to $\gamma_{rad}^{c}$.
\par
In order to explicitly incorporate the effect of the plasmon modes into our quantum master equation we start from the assumption that the atomic resonance lines up with the fundamental dipole mode, i.e. $\omega_{e}=\omega_{1}$. Any emission (both radiative and non-radiative) mediated by the $l=1$ mode will be treated through the Jaynes-Cummings terms associated to the cavity. As such the term $\hbar\Omega \left(pS_{+}+p^{\dagger}S_{-}\right)$ describes the coupling between the emitter and the fundamental $l=1$ mode ($\omega_{c}=\omega_{1}$) while $\gamma_{c}$ contains the radiative ($\gamma_{rad}$) and non-radiative ($\gamma_{abs}$) decay through the $l=1$ cavity mode. The goal is now to correlate the Jaynes-Cummings parameters ($\Omega$ and $\gamma_{c}$) to the parameters in the $l=1$ contribution of $\gamma_{tot}$. To this end we first consider the non-radiative decay of a general quantum state $\ket{\psi(t)}=c_{g}(t)\ket{g,1}+c_{e}(t)\ket{e,0}$ consisting of either the emitter in the ground state and one photon in the cavity or the emitter in the excited state and no photon in the cavity. If we only consider non-radiative decay (which for $l=1$ is mediated by $\gamma_{abs}$), the Schr\"odinger equation for $c_{g}(t)$ and $c_{e}(t)$ reduces to
\begin{align}
\frac{dc_{g}(t)}{dt}&=-i\Omega c_{e}(t)-\left(i\delta_{c}+\frac{\gamma_{abs}}{2}\right)c_{g}(t)\\
\frac{dc_{e}(t)}{dt}&=-i\Omega c_{g}(t).
\end{align}
For sufficiently small $\Omega$ one can adiabatically eliminate $c_{g}(t)$ by setting $\frac{dc_{g}(t)}{dt}=0$, which eventually results in
\begin{equation}
\frac{dc_{e}(t)}{dt}=-\frac{\Omega^{2}}{i\delta_{c}+\frac{\gamma_{abs}}{2}}c_{e}(t).
\end{equation}
The absorption mediated decay rate of the excited state emitter $\ket{e,0}$ due to the $l=1$ mode is then determined by
\begin{equation}
\frac{d|c_{e}(t)|^{2}/dt}{|c_{e}(t)|^{2}}=-\frac{\Omega^{2}\gamma_{abs}}{\delta_{c}^{2}+\left(\frac{\gamma_{abs}}{2}\right)^{2}}.
\end{equation}
The latter decay rate should now be identified with the $l=1$ contribution of $\gamma_{nrad}^{c}$, i.e.
\begin{equation}
\Omega_{1}^{2}\left(\frac{\frac{\gamma_{i}}{2}}{\left(\omega-\omega_{1}\right)^{2}+\left(\frac{\gamma_{i}}{2}\right)^{2}}\right)\equiv\frac{\Omega^{2}\gamma_{abs}}{\delta_{c}^{2}+\left(\frac{\gamma_{abs}}{2}\right)^{2}}.
\end{equation}
Since $\delta_{c}=\omega-\omega_{1}$, we can identify $\gamma_{abs}=\gamma_{i}$ and $\Omega=\frac{\Omega_{1}}{\sqrt{2}}$. The above analysis can be repeated for a general cavity decay rate $\gamma_{c}=\gamma_{rad}+\gamma_{abs}$. On resonance ($\delta_{c}=0$) the absorption part of the decay rate is then $\gamma_{abs}\left(4\Omega^{2}/\gamma_{c}^{2}\right)$ while the radiative part is $\gamma_{rad}\left(4\Omega^{2}/\gamma_{c}^{2}\right)$. At optical frequencies the Drude model parameters of gold are $\omega_{p}=14\times 10^{3}$ THz, $\epsilon_{\infty}=9$, $\gamma_{i}=100$ THz (Ref. \cite{refSN1}). For a cavity resonance around 637 nm, this implies an absorption $Q-$factor of $Q\approx 15$. For small particles the huge $\gamma_{abs}$ will usually dominate over $\gamma_{rad}$, meaning that the overall cavity decay rate $\gamma_{c}\approx\gamma_{abs}$ can be well approximated by a single $Q-$factor of $Q\approx 15$ for visible frequencies. Since the radiative decay rate $\gamma_{rad}^{c}$ can be calculated analytically for a dipole near a spherical metallic nanoparticle, $\gamma_{rad}$ is then eventually determined by $\gamma_{rad}\approx\gamma_{rad}^{c}\left(\gamma_{abs}^{2}/4\Omega^{2}\right)$. However, for the purpose of evaluating the master equation, we only need $\gamma_{c}$ which as shown is well described by a single $Q\approx 15$, i.e. $\gamma_{c}\approx\frac{\omega_{c}}{2Q}$ with $\omega_{c}\approx 2957$ THz (637 nm).
\par
Now we only need to determine the effective decay rate $\gamma_{e}$, which incorporates emission into all channels other than the $l=1$ cavity mode. Since the atom is off-resonant with the higher order modes one can neglect strong coupling effects with these modes. In that case we can adiabatically eliminate the higher order modes ($l>1$) and incorporate them into this effective decay rate (i.e. a local density of states). As we outlined before, the dipole mode is the only one contributing to the radiative emission, so apart from the intrinsic decay rate $\gamma_{d}$, the effective decay rate $\gamma_{e}$ hence contains the non-radiative part of all higher order modes ($l>1$), i.e.
\begin{equation}
\gamma_{e}=\gamma_{d}+\frac{\Omega_{1}^{2}}{2}f_{q}=\gamma_{d}+\Omega^{2}f_{q}
\end{equation}
with
\begin{equation}
\footnotesize
f_{q}=\sum_{l=2}^{\infty} \left(\frac{(2l+1)(l+1)^{2}}{12l (1+\xi)^{2l-2}}\right)\left(\frac{\gamma_{c}}{\left(\omega_{e}-\omega_{l}\right)^{2}+\left(\frac{\gamma_{c}}{2}\right)^{2}}\right)
\end{equation}
\normalsize
where we used the fact that the master equation is evaluated in a frame rotating at $\omega=\omega_{e}$ and that $\gamma_{i}=\gamma_{abs}\approx\gamma_{c}$. In a medium with relative permittivity $\epsilon_{d}$ the cavity modal volume of a spherical particle equals $V_{c}=\frac{\pi R^{3}}{\epsilon_{d}}$, such that
\begin{equation}
\Omega^{2}=\frac{9c^{3}\gamma_{0}}{2\omega_{p}^{2}R^{3}(1+\xi)^{6}}=\frac{9\pi c^{3}\gamma_{0}}{2\epsilon_{d}\omega_{p}^{2}\left(1+\xi\right)^{6}}\left(\frac{1}{V_{c}}\right).
\end{equation}
Now we have fully expressed the parameters appearing in the quantum master equation (i.e. $\Omega,\gamma_{c},\gamma_{e}$) in terms of the parameters for dipole emission near a spherical metallic nanoparticle. The only remaining parameter in the master equation is $\kappa$, which will be derived in the next section.

\subsection{Derivation of $\kappa$}

In order to derive the coupling constant between the waveguide and the cavity we note that the plasmonic mode has a dipole moment $\textbf{d}_{c}$ associated to the specific charge distribution $\rho(\textbf{r})$ of the mode, i.e.
\begin{equation}
\textbf{d}_{c}(\textbf{r}_{c})=\iiint \rho(\textbf{r})\left[\textbf{r}-\textbf{r}_{c}\right]d\textbf{r}\overset{\Delta}{=}\alpha_{p}\textbf{E}_{c}(\textbf{r}_{c}) 
\end{equation}
where we have introduced the plasmon polarizability $\alpha_{p}$ relating $\textbf{d}_{c}(\textbf{r}_{c})$ to the cavity field $\textbf{E}_{c}(\textbf{r}_{c})$ (both evaluated at the center of the antenna $\textbf{r}_{c}$). Dot-multiplying the above equation with $\overline{\textbf{E}}_{c}(\textbf{r}_{c})$ eventually results in 
\footnotesize
\begin{align}
\alpha_{p} &= \frac{\overline{\textbf{E}}_{c}(\textbf{r}_{c})\cdot \left(\iiint \rho(\textbf{r})\left[\textbf{r}-\textbf{r}_{c}\right]d\textbf{r} \right)}{\left|\textbf{E}_{c}(\textbf{r}_{c})\right|^{2}} \nonumber \\
&=\epsilon_{0}\frac{\overline{\textbf{E}}_{c}(\textbf{r}_{c})\cdot \left(\iiint \nabla\cdot \textbf{D}_{r}(\textbf{r})\left[\textbf{r}-\textbf{r}_{c}\right]d\textbf{r} \right)}{\left|\textbf{E}_{c}(\textbf{r}_{c})\right|^{2}} \nonumber \\
&=\epsilon_{0}\left(\frac{\epsilon_{d}\overline{\textbf{E}}_{c}(\textbf{r}_{c})\cdot \left(\iiint \nabla\cdot \textbf{D}_{r}(\textbf{r})\left[\textbf{r}-\textbf{r}_{c}\right]d\textbf{r} \right)}{\iiint d\textbf{r}\epsilon(\textbf{r})\left|E_{c}(\textbf{r})\right|^{2}}\frac{\left|\textbf{E}_{c}^{m}\right|^{2}}{\left|\textbf{E}_{c}(\textbf{r}_{c})\right|^{2}}\right)V_{c} \nonumber \\
&=\epsilon_{0}\alpha_{0}V_{c}
\end{align}
\normalsize
where 
\begin{equation}
V_{c}=\frac{\iiint d\textbf{r}\epsilon(\textbf{r})\left|E_{c}(\textbf{r})\right|^{2}}{\epsilon_{d}\left|\textbf{E}_{c}^{m}\right|^{2}}
\end{equation} 
is the cavity modal volume defined using the maximum cavity electric field strength $\left|\textbf{E}_{c}^{m}\right|$ (obtained in a region with relative dielectric permittivity $\epsilon_{d}$). Moreover we factored out $\epsilon_{0}$ such that the dielectric displacement is calculated using the relative permittivities ($\textbf{D}_{r}(\textbf{r}))$. The interaction between the waveguide mode and the dipole moment of the plasmonic cavity is $-\textbf{d}_{c}(\textbf{r}_{c})\cdot \textbf{E}_{wg}(\textbf{r}_{c})=-\alpha_{p}\textbf{E}_{c}(\textbf{r}_{c})\cdot \textbf{E}_{wg}(\textbf{r}_{c})$, where $\textbf{E}_{wg}(\textbf{r}_{c})$ is the strength of the evanescent modal field of the waveguide. After normalization of the electric fields associated to the waveguide mode and the plasmon mode we can then derive the coupling constant $g_{wg}$ as
\begin{align}
g_{wg}&=\frac{\alpha_{p}}{\hbar}\sqrt{\frac{L}{2\pi}}\left(\sqrt{\frac{\hbar\omega_{c}|\textbf{E}_{c}(\textbf{r}_{c})|^{2}}{2\iiint_{V_{c}} d\textbf{r}\epsilon_{0}\epsilon(\textbf{r})\left|\textbf{E}_{c}(\textbf{r})\right|^{2}}}\right)\times \nonumber \\
&\ \ \ \left(\sqrt{\frac{\hbar\omega_{c}|\textbf{E}_{wg}(\textbf{r}_{c})|^{2}}{2L\iint d\textbf{r}\epsilon_{0}\epsilon(\textbf{r})\left|\textbf{E}_{wg}(\textbf{r})\right|^{2}}}\right)\left(\textbf{u}_{c}\cdot\textbf{u}_{wg}\right)^{2}.
\end{align}
The length $L$ is an arbitrary length along the propagation direction and arises due to the transition from a discrete number of modes to a mode continuum (Ref. \cite{ref16}), but cancels out after normalization. \cite{ref18} The unit vectors $\textbf{u}_{c}$ and $\textbf{u}_{wg}$ define the polarization of the plasmon mode and waveguide mode respectively. The decay rate between the waveguide mode and the plasmonic cavity is then eventually
\begin{equation}
\kappa=\frac{4\pi g_{wg}^{2}}{c}=\frac{\alpha_{p}^{2}\omega_{c}^{2}}{2c\epsilon_{0}^{2}\epsilon_{d}\epsilon_{wg}}\left(\frac{1}{V_{c}}\right)\left(\frac{1}{A_{eff}}\right)\chi^{\kappa} ,
\end{equation}
where
\begin{equation}
A_{eff}=\frac{\iint d\textbf{r}\epsilon(\textbf{r})\left|\textbf{E}_{wg}(\textbf{r})\right|^{2}}{\epsilon_{wg}|\textbf{E}_{wg}^{m}|^{2}}
\end{equation} 
is the effective modal area of the waveguide mode, $|\textbf{E}_{wg}^{m}|$ the maximum electric field strength of the waveguide mode (obtained in a region with relative dielectric permittivity $\epsilon_{wg}$) and $\chi^{\kappa}=\frac{\left|\textbf{E}_{wg}(\textbf{r}_{c})\right|^{2}}{\left|\textbf{E}_{wg}^{m}\right|^{2}}\frac{\left|\textbf{E}_{c}(\textbf{r}_{c})\right|^{2}}{\left|\textbf{E}_{c}^{m}\right|^{2}}\left(\textbf{u}_{c}\cdot\textbf{u}_{wg}\right)^{2}$ a factor incorporating the overlap between the waveguide and cavity mode. Finally one gets
\begin{equation}
\kappa=\frac{\omega_{c}^{2}\chi^{\kappa}}{2cA_{eff}\epsilon_{d}\epsilon_{wg}}\left(\frac{\alpha_{0}^{2}V_{c}^{2}}{V_{c}}\right) =\frac{\omega_{c}^{2}\chi^{\kappa}\alpha_{0}^{2}V_{c}}{2cA_{eff}\epsilon_{d}\epsilon_{wg}}.
\end{equation}
The effective modal area $A_{eff}$ can also be expressed as a function of the effective mode index $n_{eff}=\sqrt{\epsilon_{eff}}$, $A_{eff}\approx\left(\frac{\lambda_{c}}{2n_{eff}}\right)^{2}$. \cite{refSN2} For a spherical particle in a background medium with relative permittivity $\epsilon_{d}$, the dipole moment of the fundamental plasmon mode is $\textbf{d}= 2\epsilon_{0}\epsilon_{d}V_{c}\textbf{E}=\alpha_{p}\textbf{E}$, i.e. $\alpha_{0}=2\epsilon_{d}$ (Ref. \cite{ref22}), such that
\begin{equation}
\kappa=\frac{\omega_{c}^{4}\alpha_{0}^{2}\chi^{\kappa}\epsilon_{eff}}{2\pi^{2}c^{3}\epsilon_{d}\epsilon_{wg}}V_{c}=\frac{2\chi^{\kappa}R^{3}\omega_{c}^{4}\epsilon_{eff}}{\pi c^{3}\epsilon_{wg}}.
\end{equation}

\subsection{Single photon extraction efficiency}

The rate equations for our quantum photonic platform are similar to the ones obtained in Ref. \cite{refN7}, but now with the addition of a decay rate term into the waveguide modes. In the basis $\{\ket{1}=\ket{g,0},\ket{2}=\ket{g,1},\ket{3}=\ket{e,0}\}$ we get
\begin{align}
\frac{d\rho_{11}}{dt}&=\gamma_{p}\rho_{22}+\gamma_{e}\rho_{33} \\
\frac{d\rho_{22}}{dt}&=-2\Omega\Im(\rho_{23})-\gamma_{p}\rho_{22}\\
\frac{d\rho_{33}}{dt}&=2\Omega\Im(\rho_{23})-\gamma_{e}\rho_{33}\\
\frac{d\Im(\rho_{23})}{dt}&=-\frac{\gamma_{p}+\gamma_{e}+\gamma^{*}}{2}\Im(\rho_{23})+\Omega\rho_{22}-\Omega\rho_{33}
\end{align}
and we assume the system is initially in the excited state, i.e. $\rho(t=0)=\ket{3}\bra{3}$. For a realistic nanoplasmonic cavity $\gamma_{p}$ usually exceeds 100 THz due to the low $Q-$factor, while dephasing rates at room temperature are typically on the order of a few THz, implying that $\gamma_{p}+\gamma_{e}+\gamma^{*}\approx\gamma_{p}+\gamma_{e}$ in the equation for $\Im(\rho_{23})$. After solving for $\rho_{22}(t)$, the single photon extraction efficiency into the waveguide mode is
\begin{equation}
\eta=\kappa\int_{0}^{\infty}dt\rho_{22}(t)=\frac{\kappa}{(\gamma_{e}+\gamma_{p})\left(1+\frac{\gamma_{e}\gamma_{p}}{4\Omega^{2}}\right)}.
\end{equation}


\subsubsection{Maximum single photon extraction efficiency}

As discussed in the main text, the single photon extraction efficiency reaches a maximum for a certain optimum $(\xi,R)$ combination (or equivalently for a certain $(\xi,V_{c})$). In this section we derive an approximate formula for these optimum values in order to assess how they depend on the physical parameters of the system. 
\par
Near resonance the quenching term $f_{q}$ can be approximated as
\footnotesize
\begin{align}
f_{q}&\approx\frac{4}{\gamma_{c}}\sum_{l=2}^{\infty} \left(\frac{(2l+1)(l+1)^{2}}{12l (1+\xi)^{2l-2}}\right) \\
&=\frac{4}{\gamma_{c}}\left(\frac{1}{12z}\sum_{l=1}^{\infty} z^{l}\left(2l^{2}+5l+4+\frac{1}{l}\right)-1\right)
\end{align}
\normalsize
with $z=\frac{1}{\left(1+\xi\right)^{2}}$. Since $0\leq\xi$, $z\leq 1$, allowing us to rewrite $f_{q}$ in terms of the polylogarithm $\text{Li}_{n}(z)=\sum_{l=1}^{\infty}\frac{z^{l}}{l^{n}}$, i.e.
\footnotesize
\begin{align}
f_{q}&=\frac{4}{\gamma_{c}}\left(\frac{1}{12z}\left(2\text{Li}_{-2}(z)+5\text{Li}_{-1}(z)+4\text{Li}_{0}(z)+\text{Li}_{1}(z)\right)-1\right) \\
&=\frac{4}{\gamma_{c}}\left(\frac{11}{12\left(1-z\right)^3}\left(1-z+\frac{4z^{2}}{11}-\frac{\left(1-z\right)^{3}}{11z}\ln\left(1-z\right)\right)-1\right).
\end{align}
\normalsize
Since the maximum $\eta$ is expected in the cross-over region between the quenching limited and intrinsic decay rate limited case, the optimum $\xi$ is expected to be in the range $0.1$ to $1$. For $\xi$ values in this range, the term $\propto (1-z)^{-2}$ will usually dominate over the other terms in $z$, allowing to further approximate $f_{q}$ as
\begin{equation}
f_{q}\approx\frac{4}{\gamma_{c}}\left(\frac{z(2-z)}{(1-z)^{2}}\right)=\frac{4}{\gamma_{c}}\left(\frac{1+4\xi+\xi^{2}}{\xi^{2}\left(4+4\xi+\xi^{2}\right)}\right).
\end{equation}
\par
As shown above, the optimum $\xi$ is expected to be in the range $0.1$ to $1$, allowing us to approximate $\gamma_{e}+\gamma_{p}\approx\gamma_{p}$ in the equation for $\eta$, i.e.
\begin{equation}
\eta\approx\frac{\kappa}{\left(\gamma_{c}+\kappa\right)+\left(\frac{\gamma_{d}}{4\Omega^{2}}+\frac{f_{q}}{4}\right)\left(\gamma_{c}+\kappa\right)^{2}}
\end{equation}
The optimum $\xi$ is determined by
\begin{equation}
\frac{\partial}{\partial z}\left(\frac{\gamma_{d}}{4\Omega^{2}}+\frac{f_{q}}{4}\right)=0
\end{equation}
which approximately yields
\begin{equation}
\xi_{opt}\approx\sqrt[18]{\frac{44\chi^{\kappa}Q^{2}}{\pi n_{d}}\left(\frac{\omega_{c}}{\omega_{p}}\right)^{2}\left(\frac{\alpha_{0}}{\epsilon_{d}}\right)^{2}\left(\frac{\epsilon_{eff}}{\epsilon_{wg}}\right)}-1.
\end{equation}
Using the parameter values from the main text, $\xi_{opt}\approx 0.35$ which is in excellent correspondence with the numerically obtained optimum $\xi$ of 0.38. In general $\frac{\partial \eta}{\partial V_{c}}$ yields a cubic equation in $V_{c}$. A simple estimate for the optimum cavity modal volume can however be obtained in the limit $\frac{f_{q}}{4}>\frac{\gamma_{d}}{4\Omega^{2}}$, which is a valid first approximation in the cross-over region. In this limit one gets
\begin{align}
V_{\eta}^{opt}&\approx\frac{\lambda_{c}^{3}}{Q}\left(\frac{\epsilon_{d}\epsilon_{wg}}{8\pi\chi^{\kappa}\alpha_{0}^{2}\epsilon_{eff}}\right)\left(\frac{(1+\xi_{opt})^{2}}{\sqrt{1+4\xi_{opt}+\xi_{opt}^{2}}}\right) \\
&\approx\frac{\lambda_{c}^{3}}{Q}\left(\frac{\epsilon_{d}\epsilon_{wg}}{8\pi\chi^{\kappa}\alpha_{0}^{2}\epsilon_{eff}}\right).
\end{align}
This value corresponds to an optimal radius of $R_{opt}\approx 38$ nm (for the parameter values used in the main text), which again is in good correspondence with the numerically obtained value of 47.5 nm. So our first approximation already yields a good prediction of the optimum values, allowing us to identify the scaling behaviour of $\xi_{opt}$ and $R_{opt}$ with the physical parameters of the system ($Q-$factor, etc.).

\subsection{Condition for high indistinguishability}
The indistinguishability in the bad cavity (BC) limit is given by
\begin{equation}
I=\frac{\gamma_{e}+R}{\gamma_{e}+\gamma^{*}+R}
\end{equation}
with $R=\frac{4\Omega^{2}}{\gamma_{e}+\gamma^{*}+\gamma_{p}}$. \cite{refN7} In the BC limit $\gamma_{p}/(\gamma_{e}+\gamma^{*})\gg 1$ such that
\begin{equation}
I\approx\frac{\gamma_{p}\gamma_{e}+4\Omega^{2}}{\gamma_{p}(\gamma_{e}+\gamma^{*})+4\Omega^{2}}.
\end{equation}
Since the dephasing rate at room temperature is typically orders of magnitude higher than the intrinsic decay rate, the condition to achieve near-unity $I$ boils down to
\begin{equation}
\left(\gamma_{c}+\kappa\right)\left(f_{q}\Omega^{2}+\gamma^{*}\right)\ll 4\Omega^{2}.
\end{equation}
Since the former condition holds in the BC limit, $\gamma_{e}=\gamma_{d}+f_{q}\Omega^{2}<\gamma_{p}$ needs to be satisfied at all times. This implies that $\Omega^{2}<(\gamma_{p}-\gamma_{d})/f_{q}$. On the other hand, the constraint for high $I$ requires $4\Omega^{2}$ to surpass at least $\gamma_{c}\gamma^{*}$. As such, the emitter cannot come to close to the surface to avoid a large quenching factor $f_{q}$ but simultaneously the cavity modal volume needs to be small enough such that $4\Omega^{2}$ can at least surpass $\gamma_{c}\gamma^{*}$. The required decrease in modal volume implies a concomitant decrease in $\kappa$, such that the constraint eventually becomes
\begin{equation}
\left(f_{q}\Omega^{2}+\gamma^{*}\right)\ll \frac{4\Omega^{2}}{\gamma_{c}}.
\end{equation}
By taking into account that $f_{q}$ cannot be too large (i.e. $f_{q}=\mathcal{O}(4/\gamma_{c})$), the constraint on the modal volume approximately is
\begin{equation}
V_{c}\ll V_{\eta}^{opt}\left(\frac{\alpha_{0}}{\epsilon_{d}}\right)^{2}\left(\frac{\epsilon_{eff}}{\epsilon_{wg}}\right)\left(\frac{\gamma_{0}}{\gamma^{*}}\right)Q^{2}\chi^{\kappa}.
\end{equation}
Due to the large dephasing rate $\frac{\gamma_{0}}{\gamma^{*}}\gg 1$, the required cavity modal volume to reach high $I$ is much smaller than the one required to obtain maximum single photon extraction efficiency. This imposes a fundamental trade-off for integrated nanoplasmonic systems at room temperature.

\subsection{Improving $\eta$ and $\eta I$}
Here we investigate the impact of field enhancement and increased polarizability on the value of $\eta$ and $\eta I$. These effects are modeled by respectively introducing a multiplication factor into the formulas for $\Omega$ and $\alpha_{0}$ of a spherical particle, i.e
\begin{align}
\Omega_{new}&=\zeta_{\Omega}\Omega_{sphere}=\zeta_{\Omega}\sqrt{\frac{9c^{3}\gamma_{0}}{2\omega_{p}^{2}R^{3}(1+\xi)^{6}}} \\
\kappa_{new}&=\kappa_{sphere}[\alpha_{0}\rightarrow\zeta_{\kappa}\alpha_{0}]=\frac{2\chi^{\kappa}R^{3}\omega_{c}^{4}\zeta_{\kappa}^{2}}{\pi c^{3}}.
\end{align}
From a physical point of view, the introduction of $\zeta_{\Omega}$ could be motivated if we e.g. bring two spherical particles close to each other and assure the gap between them is small enough to enhance the field in the gap (compared to the maximum field that can be achieved for a single spherical particle). For example, bringing two spherical particles close to each other can already lead to a field enhancement by a factor $\approx Q$. \cite{ref22} Bringing the particles closer together also affects the charge distribution on each of the spheres, resulting in a change of the dipole moment of the fundamental mode. Alternatively more intricate nanoplasmonic cavities could be considered, e.g. rod or bowtie antennas. It should be noted that a change in antenna dipole moment ($\zeta_{\kappa}>1$) not only increases $\kappa$ but also increases the radiative decay rate $\gamma_{rad}$ and as such changes the $Q-$factor. This effect is however neglected in our simple first assessment. While a complete electromagnetic simulation is hence required to exactly determine $\eta$ and $\eta I$, our initial calculation already hints on the potential improvements.
\begin{figure}[h]
\includegraphics[width=0.5\textwidth]{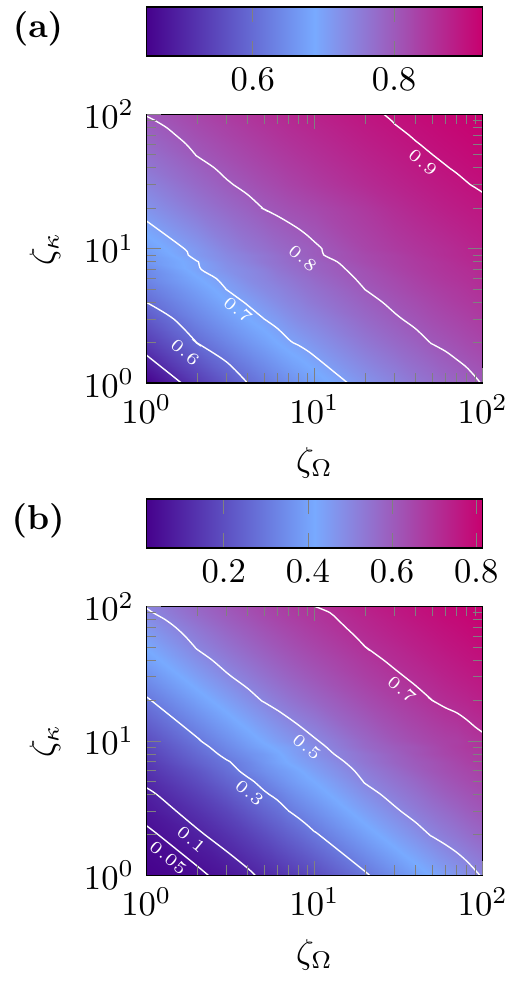}
\caption{Single photon characteristics for a nanoplasmonic cavity with $Q=15$, $\gamma_{d}=1$ GHz, $\gamma^{*}=3.5$ THz, $\epsilon_{eff}=4$, $\epsilon_{wg}=4$, $\chi^{\kappa}=1$ and the incorporation of 1000 higher order modes. \textbf{(a)} Single photon extraction efficiency $\eta$. \textbf{(b)} Efficiency-indistinguishability product $\eta I$. The combination ($\zeta_{\Omega}=1,\zeta_{\kappa}=1$) corresponds to the spherical particle case. \label{FigureSPGEAnt}}
\end{figure}
\par
The results of increased field enhancement and/or polarizability are shown in Fig. \ref{FigureSPGEAnt}, where $\eta$ (a) and $\eta I$ (b) are shown as a function of $\zeta_{\Omega}$ and $\zeta_{\kappa}$. For each $(\zeta_{\Omega},\zeta_{\kappa})$ combination, $\eta$ and $\eta I$ are evaluated at their respective optimum $(V_{c},\xi)$ values. For the reported $\zeta_{\Omega}$ and $\zeta_{\kappa}$ values it is hence possible to significantly improve on $\eta$ and $\eta I$. Enhancing the field and the polarizability by a factor 10 already allows single photon extraction efficiencies of $80\%$ while $\eta I$ increases to $50\%$ (implying that both $\eta$ and $I$ are at least $70\%$).

\end{document}